\documentclass[11pt,a4paper]{article}

\usepackage[
    a4paper,
    margin=2.5cm
]{geometry}
\usepackage{setspace}
\usepackage{indentfirst}

\usepackage[T1]{fontenc}
\usepackage[utf8]{inputenc}
\usepackage{lmodern}
\usepackage{microtype}

\usepackage{amsmath}
\usepackage{amssymb}
\usepackage{amsfonts}
\usepackage{amsthm}
\usepackage{mathtools}
\usepackage{bm}

\usepackage{graphicx}
\usepackage{booktabs}
\usepackage{array}
\usepackage{caption}
\usepackage{subcaption}
\usepackage{float}
\usepackage{placeins}

\usepackage[numbers,sort&compress]{natbib}
\usepackage{xcolor}
\usepackage{hyperref}
\hypersetup{
    colorlinks=true,
    linkcolor=blue!60!black,
    citecolor=blue!60!black,
    urlcolor=blue!60!black,
    pdfauthor={Jining Tang, Yang Huang, Hongsheng Zhang},
    pdftitle={Scalar absorption beyond geometric optics in Klein-Gordon-separable Johannsen black hole spacetimes}
}

\graphicspath{
    {figs/}
    {Numerics/figures/}
    {Derivations/}
}

\newcommand{\dd}{\mathrm{d}}
\newcommand{\ii}{\mathrm{i}}
\newcommand{\ee}{\mathrm{e}}
\newcommand{\Sigmat}{\tilde{\Sigma}}
\newcommand{\Dcal}{\mathcal{D}}

\newcommand{\siggeo}{\sigma_{\mathrm{geo}}}
\newcommand{\sigabs}{\sigma_{\mathrm{abs}}}

\DeclareMathOperator{\sinc}{sinc}
\allowdisplaybreaks

\renewcommand{\thesection}{\Roman{section}}

\makeatletter
\renewcommand{\p@subsection}{\thesection.}
\makeatother

\title{\bfseries Scalar absorption beyond geometric optics in Klein--Gordon-separable Johannsen black hole spacetimes}

\author{
    Jining Tang$^{1}$\thanks{Electronic address: aishiker1998@gmail.com}
    \quad
    Yang Huang$^{2}$\thanks{Electronic address: sps\_Huangy@ujn.edu.cn}
    \quad
    Hongsheng Zhang$^{2}$\thanks{Electronic address: sps\_zhanghs@ujn.edu.cn (contact author)}
    \\[0.4em]
    \small School of Physics and Technology, University of Jinan,\\
    \small 336 West Road of Nan Xinzhuang, Jinan, Shandong 250022, China
}

\date{\today}

\begin{document}

\maketitle

\begin{abstract}
Johannsen metric is a natural and significant generalization of the Kerr metric, representing the most general stationary, axisymmetric spacetime that preserves the Carter constant of motion. The 
theoretical status furnishes a powerful, systematic framework for strong-field tests of the no-hair theorem and for investigations of deviations from Kerr black-hole geometries.
We formulate massless scalar plane-wave absorption in a
Klein--Gordon-separable subclass of Johannsen spacetimes.  In the asymptotically flat
Johannsen metric, we impose Klein--Gordon separability, derive the separated
angular and radial equations, and build a partial wave framework for the
leading deformation sectors $A_1(r)$, $A_2(r)$, and $A_5(r)$.  The resulting
description separates deformations that change the radial size function
$X(r)$ from those that enter only the radial kinetic term. The former modify the low-frequency area law, the high-frequency null-capture cross section, and the finite-frequency absorption spectra, 
whereas a pure $A_5$ deformation leaves the leading null-capture observable unchanged while remaining detectable in wave propagation. We further examine off-axis incidence,
co-/counter-rotating contributions, and superradiant modes, where changes in
$X(r_+)$ shift the horizon angular velocity and hence the superradiant
threshold.  Our results identify finite-frequency absorption as a wave-optics diagnostic 
that can probe radial propagation sectors inaccessible to both the area law and 
null geodesic capture observables, offering a new tool for strong-field tests of black hole geometry.
\end{abstract}


\noindent\textbf{PACS/MSC/Subject classification:}
04.70.-s; 04.30.Nk; 04.50.Kd

% ======================================================================
\section{Introduction}
\label{sec:introduction}

The Kerr solution is the standard exterior geometry of an isolated rotating
black hole in general relativity~\cite{Kerr1963,Carter1968}.  Strong-field
observations now test this assumption directly through gravitational waves,
horizon-scale imaging, and related probes of compact object dynamics
\cite{Abbott2016,EHT2019,Barack2019}.  A useful theoretical strategy is
therefore to work with Kerr-type metrics that preserve enough symmetry for
observable calculations, while introducing controlled deviations that can be
connected either to phenomenology or to exact alternatives to Kerr.
We follow standard general relativity and black hole conventions as presented
in classic treatments of the subject~\cite{MisnerThorneWheeler1973,Wald1984,Chandrasekhar1983,Carroll2004}.

Parametrized rotating metrics provide such a framework.  The original
Johannsen--Psaltis construction was designed for strong-field tests of the
no-hair conjecture,
but generic deformations may introduce pathologies or lose integrability at
large dimensionless spin~\cite{JohannsenPsaltis2011,CardosoPaniRico2014}.  The
Johannsen metric used here is more suitable for analytic and numerical
absorption calculations because it is regular outside the event horizon in the allowed
parameter domain, admits a Carter-type constant, and packages deviations into
radial functions that can be compared with known black hole metrics for
special parameter choices~\cite{Johannsen2013}.  Closely related work on
parametrized Kerr-type geometries clarifies the role of asymptotic flatness,
coordinate choices, hidden symmetries, and separability in restricting
admissible deviations~\cite{CarsonYagi2020,Frolov2017}.

The distinction between geodesic separability and wave separability is central
to the present problem.  Hamilton--Jacobi separability makes photon orbits and
geometric observables tractable, but scalar absorption requires the
Klein--Gordon (KG) equation to separate in the same coordinates.  Konoplya,
Stuchl{\'i}k, and Zhidenko constructed a broad axisymmetric class admitting
both Hamilton--Jacobi and Klein--Gordon separability, while Chen and Chen
derived analogous constraints for rotating spacetimes generated through the
Newman-Janis algorithm~\cite{Konoplya2018,Chen2019}.  Shadow calculations in
general rotating spacetimes and in Johannsen-type geometries further show
that null geodesic observables probe only selected combinations of the metric
functions~\cite{Shaikh2019,LongChenWangJing2020,PerlickTsupko2022,Meng2023}.
This motivates asking which deformation sectors remain visible when one goes
beyond geometric optics.

Black-hole absorption supplies a complementary probe.  The classic scalar
wave problem ranges from low energy absorption by small black holes to
Schwarzschild absorption spectra and Hawking emission greybody factors
\cite{Unruh1976,Page1976,Sanchez1978}.  In Kerr, the separable wave equations,
superradiant amplification, and partial wave absorption formalism were developed
through the Teukolsky--Starobinsky--Press framework and later numerical
studies~\cite{Teukolsky1973,StarobinskyChurilov1973,TeukolskyPress1974,Futterman1988,Macedo2013}.
Subsequent work extended the comparison to electromagnetic and gravitational
waves, incidence-angle dependence, and co-/counter-rotating mode
decompositions~\cite{Dolan2008,LeiteDolanCrispino2017,LeiteDolanCrispino2019}.

Different frequency regimes isolate different aspects of the geometry.  In the
low-frequency regime, the massless scalar absorption cross section approaches the
horizon area for broad classes of stationary black holes
\cite{DasGibbonsMathur1997,Higuchi2001,Higuchi2002}.  In the high-frequency regime, the
cross section approaches the geometric capture area, with oscillatory
corrections described by complex angular momentum, Regge pole, and sinc
approximations tied to unstable null orbits
\cite{DecaniniFolacciJensen2003,Decanini2011,DecaniniFine2011,Benone2018,OuldElHadj2025}.
Finite-frequency spectra occupy the intermediate regime in which the full
radial wave barrier matters.  This is already visible in Kerr-Newman,
tidal-charge, regular black-hole, conformally related, and dilatonic
backgrounds, where partial-wave absorption and scattering can distinguish
structures that are degenerate in one of the limiting regimes
\cite{LeiteBenoneCrispino2017,MacedoCrispino2014,Oliveira2021,LiMiao2022,DePaula2024,HuangZhang2020Dilatonic}.
Full-wave studies of charged dilatonic black holes, for example, combine
partial-wave and Regge-pole methods with classical-geodesic and glory
approximations~\cite{HuangZhang2020Dilatonic}.  More recently, scalar
absorption and scattering by Frolov regular black holes showed that Frolov,
Reissner--Nordstr{\"o}m, and Hayward spectra can become nearly degenerate
when the relevant critical or glory impact parameters are
matched~\cite{TangHuangZhang2026Frolov}.  These results motivate the present
question in a rotating parametrized spacetime: which deformation sectors
remain visible to finite-frequency waves after the low-frequency area law and
high-frequency null-capture limit have been fixed?

In this work we formulate the finite-frequency massless scalar plane-wave
absorption problem for a Klein--Gordon-separable Johannsen subclass and use it to compare
the leading $A_1(r)$, $A_2(r)$, and $A_5(r)$ deformation sectors.  The central
question is whether finite-frequency absorption can probe radial propagation
sectors that are invisible to both the low-frequency area law and the
high-frequency null capture limit.  The paper is organized
as follows.  Section~\ref{sec:kg_separable_johannsen} defines the metric
subclass and the separability restrictions.  Section~\ref{sec:absorption_formalism}
derives the separated scalar equations and the radial boundary-value problem.
Section~\ref{sec:limiting_behavior} discusses the absorption cross section and
its low- and high-frequency limits.  Section~\ref{sec:finite_frequency_results}
presents the finite-frequency spectra, incidence-angle dependence, and
superradiant mode behavior.  Section~\ref{sec:discussion} summarizes the
physical interpretation and possible extensions.  Throughout the paper we use
geometrized units $G=c=\hbar=1$ and the metric signature $(-,+,+,+)$.

% ======================================================================
\section{Klein--Gordon-Separable Johannsen Spacetimes}
\label{sec:kg_separable_johannsen}

The Johannsen metric is a parametrized deformation of Kerr designed to keep
stationarity, axisymmetry, asymptotic flatness, and three independent
geodesic constants of motion while introducing strong field deviations from
the Kerr geometry~\cite{Johannsen2013}.  In this section we keep the
logical order explicit.  We first display the asymptotically flat Johannsen
geometry, then impose Klein--Gordon separability, and only afterwards discuss
the additional weak field restrictions used in the canonical parametrization.
This separation of assumptions is important because the KG-separable subclass
and the parametrized post-Newtonian (PPN) compatible subclass need not
coincide a priori.
In Boyer--Lindquist coordinates the Kerr seed is
\begin{align}
    ds^2_{\rm K}
    &=
    -\left(1-\frac{2Mr}{\Sigma}\right)dt^2
    -\frac{4Mar\sin^2\theta}{\Sigma}dt\,d\phi
    +\frac{\Sigma}{\Delta}dr^2+\Sigma d\theta^2
    \nonumber\\
    &\quad
    +\frac{\left[(r^2+a^2)^2-a^2\Delta\sin^2\theta\right]\sin^2\theta}
           {\Sigma}d\phi^2 ,
    \label{eq:kerr_seed_metric}
\end{align}
where
\begin{equation}
    \Delta=r^2-2Mr+a^2,\qquad
    \Sigma=r^2+a^2\cos^2\theta .
    \label{eq:delta_sigma_def}
\end{equation}
Here $M$ is the black hole mass and
$a=J/M$ is the Kerr spin parameter, i.e., the specific angular momentum.  The
dimensionless spin is $a/M=J/M^2$.
Johannsen's starting point is a deformation of the contravariant Kerr
operator by radial functions $A_1,A_2,A_5,f$ and angular functions
$A_3,A_4,A_6,g$,
\begin{align}
    \Sigmat\,g^{\alpha\beta}
    \frac{\partial}{\partial x^\alpha}
    \frac{\partial}{\partial x^\beta}
    &=
    -\frac{\left[(r^2+a^2)A_1(r)\frac{\partial}{\partial t}
    +aA_2(r)\frac{\partial}{\partial \phi}\right]^2}{\Delta}
    +\frac{\left[A_3(\theta)\frac{\partial}{\partial \phi}
    +a\sin^2\theta A_4(\theta)\frac{\partial}{\partial t}\right]^2}
    {\sin^2\theta}
    \nonumber\\
    &\quad
    +\Delta A_5(r)\frac{\partial^2}{\partial r^2}
    +A_6(\theta)\frac{\partial^2}{\partial \theta^2},
    \label{eq:johannsen_contravariant_ansatz}
\end{align}
with
\begin{equation}
    \Sigmat=\Sigma+f(r)+g(\theta).
    \label{eq:tilde_sigma_general}
\end{equation}
The separable ansatz is imposed at the Hamilton--Jacobi level.  For a test
particle of rest mass $m_0$,
\begin{equation}
    -\frac{\partial S}{\partial\tau}
    =
    \frac{1}{2}g^{\alpha\beta}
    \frac{\partial S}{\partial x^\alpha}
    \frac{\partial S}{\partial x^\beta},
    \label{eq:hj_equation_short}
\end{equation}
and
\begin{equation}
    S=\frac{1}{2}m_0^2\tau-Et+L_z\phi+S_r(r)+S_\theta(\theta).
    \label{eq:hj_ansatz_short}
\end{equation}
Substitution gives
\begin{align}
    -m_0^2
    &=
    -\frac{\left[-(r^2+a^2)A_1(r)E+aA_2(r)L_z\right]^2}
    {\Delta\Sigmat}
    +\frac{\left[A_3(\theta)L_z
    -aA_4(\theta)E\sin^2\theta\right]^2}
    {\Sigmat\sin^2\theta}
    \nonumber\\
    &\quad
    +\frac{\Delta A_5(r)}{\Sigmat}
    \left(\frac{\dd S_r}{\dd r}\right)^2
    +\frac{A_6(\theta)}{\Sigmat}
    \left(\frac{\dd S_\theta}{\dd\theta}\right)^2 .
    \label{eq:hj_substitution_johannsen}
\end{align}
After multiplying by $\Sigmat$ the radial and angular terms separate.  The
separation constant $C$ may be written equivalently as
\begin{align}
    C
    &=
    \frac{\left[-(r^2+a^2)A_1(r)E+aA_2(r)L_z\right]^2}{\Delta}
    -m_0^2\left[r^2+f(r)\right]
    -\Delta A_5(r)\left(\frac{\dd S_r}{\dd r}\right)^2,
    \label{eq:C_radial_johannsen}\\
    C
    &=
    \frac{\left[A_3(\theta)L_z-aA_4(\theta)E\sin^2\theta\right]^2}
    {\sin^2\theta}
    +m_0^2\left[a^2\cos^2\theta+g(\theta)\right]
    +A_6(\theta)\left(\frac{\dd S_\theta}{\dd\theta}\right)^2.
    \label{eq:C_angular_johannsen}
\end{align}
The Carter-type constant is
\begin{equation}
    Q=C-(L_z-aE)^2 .
    \label{eq:carter_relation_short}
\end{equation}
Following the convention discussed by de Felice and Preti~\cite{DeFelicePreti1999},
we also define
\begin{equation}
    \Lambda=C+2aEL_z-m_0^2a^2 .
    \label{eq:lambda_def_geodesic}
\end{equation}
This quantity reduces to the ordinary squared angular momentum in the
asymptotically flat, nonrotating limit.  We use the geodesic construction
only to identify the metric class; the absorption calculation below is based
on the Klein--Gordon equation.

The radial deviation functions are expanded as
\begin{equation}
    A_i(r)=\sum_{n=0}^{\infty}\alpha_{in}
    \left(\frac{M}{r}\right)^n,\qquad i=1,2,5,
    \label{eq:Ai_expansion_general}
\end{equation}
while
\begin{equation}
    f(r)=\sum_{n=0}^{\infty}\epsilon_n\frac{M^n}{r^{n-2}},
    \qquad
    g(\theta)=M^2\sum_{k,l=0}^{\infty}
    \gamma_{kl}\sin^k\theta\cos^l\theta,
    \qquad \gamma_{00}=0 .
    \label{eq:f_g_expansion_general}
\end{equation}
Asymptotic flatness fixes
\begin{equation}
    \alpha_{10}=\alpha_{20}=\alpha_{50}=1,\qquad
    \epsilon_0=\epsilon_1=0,\qquad
    A_3(\theta)=A_4(\theta)=A_6(\theta)=1 .
    \label{eq:asymptotic_flatness_conditions}
\end{equation}
If the parameters $M$ and $a$ are further required to coincide with the
physical mass and Kerr spin parameter in the canonical Johannsen
normalization, then
\begin{equation}
    \alpha_{11}=\alpha_{21}=\alpha_{51}=0 .
    \label{eq:mass_spin_normalization}
\end{equation}
With these restrictions the nonzero covariant components take the compact
form
\begin{equation}
    \Dcal(r,\theta)
    =(r^2+a^2)A_1(r)-a^2A_2(r)\sin^2\theta .
    \label{eq:metric_denominator_D}
\end{equation}
\begin{align}
    g_{tt}
    &=
    -\frac{\Sigmat\left[\Delta-a^2A_2(r)^2\sin^2\theta\right]}
    {\Dcal^2},
    \label{eq:johannsen_gtt_af}\\
    g_{t\phi}
    &=
    -\frac{
    a\left[(r^2+a^2)A_1(r)A_2(r)-\Delta\right]
    \Sigmat\sin^2\theta}
    {\Dcal^2},
    \label{eq:johannsen_gtphi_af}\\
    g_{rr}
    &=
    \frac{\Sigmat}{\Delta A_5(r)},
    \qquad
    g_{\theta\theta}=\Sigmat,
    \label{eq:johannsen_grr_gthth_af}\\
    g_{\phi\phi}
    &=
    \frac{\Sigmat\sin^2\theta
    \left[(r^2+a^2)^2A_1(r)^2-a^2\Delta\sin^2\theta\right]}
    {\Dcal^2}.
    \label{eq:johannsen_gphiphi_af}
\end{align}
Equations~\eqref{eq:johannsen_gtt_af}--\eqref{eq:johannsen_gphiphi_af}
reduce to Kerr when all deviations vanish.

We next impose separability of a massive scalar field.  Let
\begin{equation}
    \left(\Box-\mu^2\right)\Phi=0,\qquad
    \Phi=e^{-\ii\omega t}e^{\ii m\phi}\psi(r,\theta),
    \label{eq:massive_kg_general}
\end{equation}
where $\mu$ is the scalar field mass.  Substituting
Eq.~\eqref{eq:massive_kg_general} into the asymptotically flat Johannsen
metric and multiplying the resulting equation by $\Sigmat^2/\Dcal$ gives
\begin{align}
    &\sqrt{A_5}\frac{\partial}{\partial r}
    \left[
    \Delta\sqrt{A_5}\,
    \frac{\Sigmat}{\Dcal}\frac{\partial\psi}{\partial r}
    \right]
    +\frac{1}{\sin\theta}
    \frac{\partial}{\partial\theta}
    \left[
    \sin\theta\,
    \frac{\Sigmat}{\Dcal}\frac{\partial\psi}{\partial\theta}
    \right]
    \nonumber\\
    &\quad
    +\frac{\Sigmat}{\Dcal\,\Delta}
    \left\{
    \left[\omega(r^2+a^2)A_1(r)-amA_2(r)\right]^2
    -\Delta\left(a\omega\sin\theta-\frac{m}{\sin\theta}\right)^2
    \right\}\psi
    -\mu^2\frac{\Sigmat^2}{\Dcal}\psi=0 .
    \label{eq:kg_full_before_Y}
\end{align}
This form motivates the definition
\begin{equation}
    Y(r,\theta)\equiv\frac{\Sigmat}{\Dcal}.
    \label{eq:Y_factor_def}
\end{equation}
It is also convenient to set
\begin{equation}
    \mathcal{N}(r,\theta)
    =
    \left[\omega(r^2+a^2)A_1(r)-amA_2(r)\right]^2
    -\Delta\left(a\omega\sin\theta-\frac{m}{\sin\theta}\right)^2 .
    \label{eq:kg_N_part}
\end{equation}
Then Eq.~\eqref{eq:kg_full_before_Y} becomes
\begin{equation}
    \sqrt{A_5}\partial_r
    \left(\Delta\sqrt{A_5}\,Y\partial_r\psi\right)
    +\frac{1}{\sin\theta}\partial_\theta
    \left(\sin\theta\,Y\partial_\theta\psi\right)
    +Y\frac{\mathcal{N}}{\Delta}\psi
    -\mu^2\Sigmat Y\psi=0 .
    \label{eq:kg_compact_structure}
\end{equation}
After division by $Y\psi$, the mass term is $-\mu^2\Sigmat$ and remains
additively separable.  The nontrivial obstruction is therefore the factor
$Y$.  With $y=\cos\theta$,
\begin{equation}
    Y(r,y)=
    \frac{r^2+f(r)+a^2y^2+g(y)}
    {(r^2+a^2)A_1(r)-a^2A_2(r)+a^2A_2(r)y^2}.
    \label{eq:Y_factor_y}
\end{equation}
Separation requires $Y(r,y)=Y_r(r)Y_y(y)$, equivalently
$\frac{\partial^2}{\partial r\,\partial y}\ln Y=0$.  For nonzero spin
parameter this condition forces $g(y)$ to be independent of $y$, so
$g(\theta)=g_0$ is a constant.  Since $\Sigmat$ depends on $f(r)$ and
$g(\theta)$ only through the combination $f(r)+g(\theta)$, this constant can
be absorbed into the constant part of $f(r)$.  With the Johannsen convention
$\gamma_{00}=0$, we choose the representative
\begin{equation}
    g(\theta)=0,
    \label{eq:kg_gtheta_zero}
\end{equation}
and the remaining radial functions must satisfy
\begin{equation}
    (r^2+a^2)A_1(r)
    =
    \left[r^2+a^2+f(r)\right]A_2(r).
    \label{eq:kg_separability_constraint}
\end{equation}
Equivalently,
\begin{equation}
    f(r)=\frac{(r^2+a^2)A_1(r)}{A_2(r)}-(r^2+a^2).
    \label{eq:f_from_a1_a2}
\end{equation}
Combining the KG separability constraint in
Eq.~\eqref{eq:kg_separability_constraint} with the power series
definitions of $A_1$, $A_2$, and $f$ gives, for the first two nontrivial
coefficients,
\begin{equation}
    \alpha_{12}=\epsilon_2+\alpha_{22},
    \qquad
    \alpha_{13}=\epsilon_3+\alpha_{23}.
    \label{eq:kg_coefficient_constraints}
\end{equation}

This KG-separable Johannsen subclass is a special form of the separable
three function metric of Konoplya, Stuchl{\'i}k, and Zhidenko~\cite{Konoplya2018}.
The comparison is useful mainly as a structural check: it shows which
Johannsen functions control the redshift sector, the effective radial size,
and the radial kinetic term in a general separable rotating metric.  If that
metric is written in terms of $R_M(r)$, $R_\Sigma(r)$, and $R_B(r)$, the
corresponding Johannsen functions obey
\begin{align}
    A_2(r)
    &=
    \left[
    \frac{\Delta}
    {r^2R_\Sigma(r)+a^2-rR_M(r)}
    \right]^{1/2},
    \qquad
    A_1(r)=
    \frac{r^2R_\Sigma(r)+a^2}{r^2+a^2}A_2(r),
    \label{eq:a1_a2_from_konoplya}\\
    A_5(r)
    &=
    \frac{1}{R_B(r)^2A_2(r)^2},\qquad
    f(r)=r^2\left[R_\Sigma(r)-1\right].
    \label{eq:a5_from_konoplya}
\end{align}
Large radius expansions of these relations connect the Johannsen coefficients
to several exact or phenomenological rotating metrics.  The following
identifications are matches at the level of coefficients in the asymptotic
expansion; when a target solution has long range charge terms, the mapping should not be
read as membership in the final working subclass.  For the Kerr--Newman
solution~\cite{Newman1965},
\begin{equation}
    \alpha_{12}=\alpha_{22}=-\frac{Q^2}{2M^2},\qquad
    \alpha_{13}=\alpha_{23}=-\frac{Q^2}{M^2},\qquad
    \alpha_{52}=\frac{Q^2}{M^2},
    \label{eq:kn_large_r_mapping}
\end{equation}
where $Q$ denotes the electric charge.  For the modified Kerr metric of
Ref.~\cite{KonoplyaZhidenko2016},
\begin{equation}
    \alpha_{13}=\alpha_{23}=\frac{\eta}{2M^3},\qquad
    \alpha_{53}=-\frac{\eta}{M^3},\qquad
    \alpha_{22}=\alpha_{52}=0 .
    \label{eq:modified_kerr_large_r_mapping}
\end{equation}
For the Kerr--Sen solution~\cite{Sen1992}, the leading nonzero coefficients
are
\begin{align}
    \epsilon_1=\frac{2b}{M},\qquad
    \alpha_{11}=\frac{b}{M},\qquad
    \alpha_{21}=-\frac{b}{M},\qquad
    \alpha_{51}=\frac{2b}{M},
    \label{eq:kerr_sen_large_r_mapping_1}\\
    \alpha_{12}=-\frac{2b}{M}-\frac{b^2}{2M^2},\qquad
    \alpha_{22}=-\frac{2b}{M}+\frac{3b^2}{2M^2},\qquad
    \alpha_{52}=\frac{4b}{M}.
    \label{eq:kerr_sen_large_r_mapping_2}
\end{align}
The $1/r$ coefficients in Eq.~\eqref{eq:kerr_sen_large_r_mapping_1} show that
Kerr--Sen is outside the canonical asymptotically normalized Johannsen
subclass unless an additional coordinate and mass parameter redefinition is
performed.  These examples should therefore be read as asymptotic coefficient
identifications.  A finite Johannsen truncation is a controlled
phenomenological approximation to the large radius behavior of a target
solution, not a global equality of the two metrics.

More generally, the Johannsen metric is not a solution of a specified set of
field equations for arbitrary deviation functions.  Its usefulness is instead
phenomenological: with suitable choices of the deviation functions, it can be
related to several known four dimensional black hole geometries discussed in
alternative gravity or beyond-Kerr tests, including modified gravity bumpy
Kerr metrics, slowly rotating dynamical Chern--Simons black holes, static
Einstein--Dilaton--Gauss--Bonnet black holes, and rotating braneworld black
holes~\cite{Johannsen2013,VigelandYunesStein2011,YunesStein2011,YunesPretorius2009,AlievGumrukcuoglu2005}.
We use these mappings only as interpretive guidance for the deformation
sectors, not as an assumption that the truncated metric solves any particular
modified field equation.

The usual PPN restrictions can now be imposed as an additional weak field
requirement.  They are not used to derive the KG condition above; instead,
they select the weak field normalization of the already separable metric.  In
the standard PPN framework~\cite{WillNordtvedt1972,Will2014,Will2018}, the
spherically symmetric weak field reference metric may be written as
\begin{equation}
    ds^2_{\rm PPN}
    =
    -A_{\rm PPN}(r)dt^2+B_{\rm PPN}(r)dr^2+r^2d\Omega^2,
    \label{eq:ppn_metric_reference}
\end{equation}
with
\begin{align}
    A_{\rm PPN}(r)
    &=
    1-\frac{2M}{r}
    +2(\beta_{\rm PPN}-\gamma_{\rm PPN})\frac{M^2}{r^2},
    \nonumber\\
    B_{\rm PPN}(r)
    &=
    1+2\gamma_{\rm PPN}\frac{M}{r}.
    \label{eq:ppn_functions_reference}
\end{align}
General relativity corresponds to
$\beta_{\rm PPN}=\gamma_{\rm PPN}=1$.  Since KG separability has already fixed
$g(\theta)=0$, the large-$r$ expansion of the asymptotically flat
KG-separable Johannsen metric gives
\begin{equation}
    ds^2=
    -\left[
    1-\frac{2M}{r}
    -\frac{M^2(2\alpha_{12}-\epsilon_2)}{r^2}
    +O(r^{-3})
    \right]dt^2
    +\left[1+\frac{2M}{r}+O(r^{-2})\right]dr^2
    +r^2d\Omega^2 .
    \label{eq:johannsen_large_r_metric}
\end{equation}
Comparison with Eq.~\eqref{eq:ppn_metric_reference} yields
\begin{align}
    2(\beta_{\rm PPN}-\gamma_{\rm PPN})
    &=
    \epsilon_2-2\alpha_{12},
    \nonumber\\
    \gamma_{\rm PPN}
    &=1,
    \label{eq:ppn_parameter_relations}
\end{align}
or equivalently
\begin{equation}
    \beta_{\rm PPN}-1
    =
    \frac{1}{2}
    \left(\epsilon_2-2\alpha_{12}\right).
    \label{eq:beta_ppn_relation}
\end{equation}
Solar-System tests constrain this quantity to be very small~\cite{Will2014,Will2018},
which we denote schematically as
\begin{equation}
    |\beta_{\rm PPN}-1|\ll1 .
    \label{eq:beta_ppn_bound}
\end{equation}
Without a tuned cancellation between the two weak field coefficients, the
PPN-compatible choice is
\begin{equation}
    \epsilon_2=\alpha_{12}=0 .
    \label{eq:ppn_simplifying_choice}
\end{equation}
Combining Eq.~\eqref{eq:ppn_simplifying_choice} with
Eq.~\eqref{eq:kg_coefficient_constraints} gives $\alpha_{22}=0$.  The
PPN-compatible truncation used below is therefore not the most general
KG-separable Johannsen subclass; it is the minimal weak field normalized
representative used for the finite-frequency calculations.  The working
geometry used in the rest of this paper is therefore
Eqs.~\eqref{eq:johannsen_gtt_af}--\eqref{eq:johannsen_gphiphi_af} with
$g(\theta)=0$ and
\begin{equation}
    \Sigmat=r^2+a^2\cos^2\theta+f(r),
    \label{eq:tilde_sigma_working}
\end{equation}
and
\begin{equation}
    A_1(r)=1+\alpha_{13}\left(\frac{M}{r}\right)^3+\cdots,\qquad
    A_2(r)=1+\alpha_{23}\left(\frac{M}{r}\right)^3+\cdots,
    \label{eq:a1_a2_expansion}
\end{equation}
\begin{equation}
    A_5(r)=1+\alpha_{52}\left(\frac{M}{r}\right)^2+\cdots,\qquad
    f(r)=\frac{(r^2+a^2)A_1(r)}{A_2(r)}-(r^2+a^2).
    \label{eq:a5_f_working_expansion}
\end{equation}
The same exterior regularity requirements used in the Johannsen
parametrization constrain the leading coefficients.  Requiring regularity
outside the event horizon, absence of determinant singularities, preservation
of the Lorentzian signature, and no closed timelike curves gives~\cite{Johannsen2013}
\begin{align}
    \alpha_{52}
    &>
    -\frac{\left(M+\sqrt{M^2-a^2}\right)^2}{M^2},
    \nonumber\\
    \epsilon_3
    &>
    -\frac{\left(M+\sqrt{M^2-a^2}\right)^3}{M^3},
    \nonumber\\
    \alpha_{13}
    &>
    -\frac{\left(M+\sqrt{M^2-a^2}\right)^3}{M^3},
    \nonumber\\
    \alpha_{22}
    &>
    -\frac{\left(M+\sqrt{M^2-a^2}\right)^2}{M^2}.
    \label{eq:johannsen_regular_domain}
\end{align}
For the weak field normalized KG-separable subclass, this bound is read
together with $\alpha_{22}=0$ and
$\epsilon_3=\alpha_{13}-\alpha_{23}$.  Since the working parametrization uses
$\alpha_{23}$ rather than the standard leading $\alpha_{22}$ as an independent
input, the analytic bounds in Eq.~\eqref{eq:johannsen_regular_domain} are
supplemented in the numerical analysis by an explicit exterior-domain check of
$\Dcal(r,\theta)$, $\Sigmat$, $A_5(r)$, and $g_{\phi\phi}$ for every parameter
set used below.  All parameter choices used in the following calculations and
discussion are therefore taken inside this exterior regularity domain, within
the KG-separable Johannsen subclass after imposing the PPN weak field
normalization above.

% ======================================================================
\section{Scalar Wave Separation and Absorption Formalism}
\label{sec:absorption_formalism}

We now turn from the metric construction to the plane-wave absorption problem.
In Sec.~\ref{sec:kg_separable_johannsen} the scalar mass $\mu$ was retained
only to state the KG separability condition at the metric level.  All
numerical absorption calculations below set $\mu=0$, so the asymptotic states
are massless scalar plane waves and can be compared directly with Kerr
results.  The governing equation is
\begin{equation}
    \Box\Phi
    =
    \frac{1}{\sqrt{-g}}
    \frac{\partial}{\partial x^\alpha}
    \left(
    \sqrt{-g}\,g^{\alpha\beta}
    \frac{\partial\Phi}{\partial x^\beta}
    \right)
    =0 .
    \label{eq:kg_equation}
\end{equation}
We solve the radial scattering boundary-value problem in order to extract the
transmission factors and absorption cross sections.  Differential scattering
cross sections are not considered in this work.
Equation~\eqref{eq:kg_compact_structure}, together with the separability
condition in Eq.~\eqref{eq:kg_separability_constraint}, is then continued
with the monochromatic mode decomposition
\begin{equation}
    \Phi(t,r,\theta,\phi)
    =
    \ee^{-\ii\omega t}\ee^{\ii m\phi}R_{lm}(r)S_{lm}(\theta).
    \label{eq:separated_ansatz}
\end{equation}
Here $\omega$ is the real frequency measured with respect to the asymptotic
time coordinate and $m$ is the azimuthal number associated with the axial
Killing field.  For each fixed $(\omega,m)$ the integer $l\geq |m|$ labels the
regular angular eigenfunction that enters the partial wave expansion of an
incident plane wave.
It is useful to introduce the radial size function
\begin{equation}
    X(r)\equiv r^2+a^2+f(r)
    =
    \frac{(r^2+a^2)A_1(r)}{A_2(r)}
    \label{eq:radial_size_function}
\end{equation}
and the radial wave factor
\begin{equation}
    W(r)=A_2(r)\left[\omega X(r)-am\right].
    \label{eq:w_function_main}
\end{equation}
The symbol $X(r)$ is only a background function and should not be confused
with the radial wave amplitude $R_{lm}(r)$.

Substitution of Eq.~\eqref{eq:separated_ansatz} separates the angular part
into the same scalar spheroidal harmonic eigenvalue problem that appears in
the Kerr scalar wave problem~\cite{Teukolsky1973,Flammer1957,Futterman1988},
\begin{equation}
    \frac{1}{\sin\theta}\frac{\dd}{\dd\theta}
    \left(\sin\theta\frac{\dd S_{lm}}{\dd\theta}\right)
    +\left[
        a^2\omega^2\cos^2\theta
        -\frac{m^2}{\sin^2\theta}
        +\lambda_{lm}(a\omega)
    \right]S_{lm}=0 .
    \label{eq:angular_equation_main}
\end{equation}
The eigenvalue $\lambda_{lm}(a\omega)$ reduces to $l(l+1)$ in the spherical
limit $a\omega\rightarrow0$.  Since the angular equation is unchanged from
Kerr, the standard scalar spheroidal basis and its selection rules can be used
without modification.  All Johannsen information relevant for the absorption
amplitudes is therefore carried by the radial equation.
With the same eigenvalue convention, the radial equation is
\begin{equation}
    A_2\sqrt{A_5}\frac{\dd}{\dd r}
    \left(
        \frac{\Delta\sqrt{A_5}}{A_2}\frac{\dd R_{lm}}{\dd r}
    \right)
    +
    \left[
        \frac{W(r)^2}{\Delta}
        -\lambda_{lm}(a\omega)-a^2\omega^2+2am\omega
    \right]R_{lm}=0,
    \label{eq:radial_equation_main}
\end{equation}
Equivalently, after defining
\begin{equation}
    \Lambda_{lm}
    \equiv
    \lambda_{lm}(a\omega)+a^2\omega^2-2am\omega ,
    \label{eq:radial_lambda_def}
\end{equation}
it can be written as
\begin{equation}
    A_2\sqrt{A_5}\frac{\dd}{\dd r}
    \left(
        \frac{\Delta\sqrt{A_5}}{A_2}\frac{\dd R_{lm}}{\dd r}
    \right)
    +
    \left[
        \frac{W(r)^2}{\Delta}
        -\Lambda_{lm}
    \right]R_{lm}=0 .
    \label{eq:radial_equation_lambda}
\end{equation}
The Kerr limit is recovered by setting $A_1=A_2=A_5=1$ and $f=0$, for which
$X=r^2+a^2$ and $W=\omega(r^2+a^2)-am$.

The structure of Eq.~\eqref{eq:radial_equation_lambda} makes the role of the
three leading deformation sectors transparent.  The functions $A_1$ and
$A_2$ determine $X(r)$ through Eq.~\eqref{eq:radial_size_function}, so
they change the algebraic wave factor $W(r)$, the horizon angular velocity,
and the turning point structure of the high-frequency problem.  By contrast,
$A_5$ enters only through the radial kinetic operator and the tortoise map.
It therefore leaves the angular equation and, for a pure $A_5$ deformation,
the geometric capture boundary unchanged, while still modifying the
finite-frequency radial propagation.

To put the radial equation into a Schr\"odinger-like wave form we define
\begin{equation}
    \frac{\dd r_*}{\dd r}
    =
    \frac{A_2(r)X(r)}{\Delta(r)\sqrt{A_5(r)}} ,
    \qquad
    U_{lm}(r)=\sqrt{X(r)}\,R_{lm}(r),
    \label{eq:tortoise_and_u_main}
\end{equation}
and
\begin{equation}
    \Xi(r)\equiv\frac{\dd r}{\dd r_*}
    =
    \frac{\Delta(r)\sqrt{A_5(r)}}{A_2(r)X(r)} .
    \label{eq:xi_main}
\end{equation}
The radial equation becomes
\begin{equation}
    \frac{\dd^2U_{lm}}{\dd r_*^2}
    +
    \left[
        \left(\omega-\frac{ma}{X(r)}\right)^2
        -V^J_{lm}(r)
    \right]U_{lm}=0 ,
    \label{eq:schrodinger_form_main}
\end{equation}
where the Johannsen scalar wave potential is
\begin{equation}
    V^J_{lm}(r)
    =
    \frac{\Delta\,\Lambda_{lm}}{A_2(r)^2X(r)^2}
    +
    \frac{\Xi(r)}{2X(r)}
    \frac{\dd}{\dd r}\!\left(\Xi(r)\frac{\dd X}{\dd r}\right)
    -
    \frac{\Xi(r)^2}{4}
    \left(\frac{1}{X(r)}\frac{\dd X}{\dd r}\right)^2 .
    \label{eq:johannsen_potential_main}
\end{equation}
This form keeps the frame dragging frequency
$\omega-ma/X(r)$ outside the potential.  The potential itself vanishes at
both the event horizon and spatial infinity under the asymptotic flatness and
regularity assumptions used here.  Thus the two ends of the exterior region
are asymptotic wave regions, while the finite radial domain between them
acts as the effective barrier, as in the standard black hole absorption
construction~\cite{Futterman1988}.

\begin{figure}[!t]
    \centering
    \includegraphics[width=0.62\textwidth]{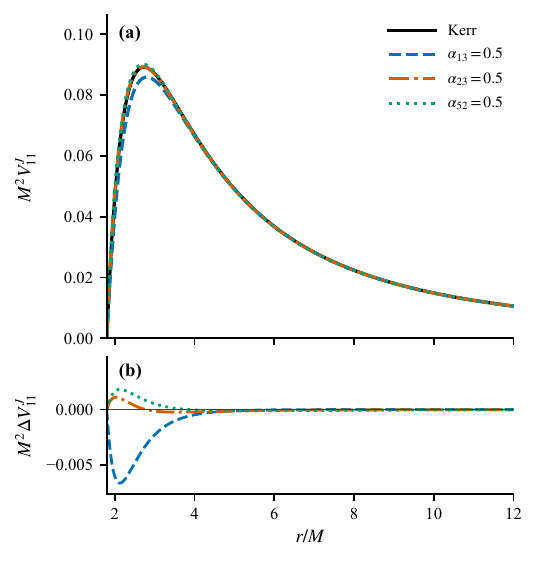}
    \caption{
        Dimensionless scalar wave effective potential $M^2V^J_{11}$ for
        representative leading Johannsen deformations.  The upper panel
        compares Kerr with the single parameter cases
        $\alpha_{13}=0.5$, $\alpha_{23}=0.5$, and $\alpha_{52}=0.5$ at
        $a/M=0.6$ and $M\omega=0.3$.  The lower panel shows the
        corresponding difference relative to Kerr, making the sector dependent
        radial response visible.  This potential is not a coordinate invariant
        observable; it is used only as a diagnostic of the radial wave barrier
        in the scattering form of Eq.~\eqref{eq:schrodinger_form_main}.
    }
    \label{fig:leading_effective_potential}
\end{figure}
The potential plot in Fig.~\ref{fig:leading_effective_potential} gives a
direct radial view of Eq.~\eqref{eq:johannsen_potential_main}.  The
deformation does not merely rescale the Kerr barrier.  In the representative
$(l,m)=(1,1)$ mode, the $A_1$ sector gives the largest change near the inner
side of the barrier, while the $A_2$ and $A_5$ sectors produce smaller but
distinct radial responses.  All deviations decay rapidly at large radius, as
required by asymptotic flatness.  The $A_5$ curve nearly overlaps with Kerr in
the upper panel, but the lower panel resolves a nonzero residual.  Thus, even
though $A_5$ drops out of the angular equation and of the leading null capture
boundary, it still changes the finite-frequency radial propagation through the
kinetic operator and the tortoise map.  For comparison with the actual
radial wave coordinate, Appendix~\ref{app:numerical_convergence}
also displays the same potential curves as functions of the tortoise
coordinate in Fig.~\ref{fig:leading_effective_potential_rstar}; the near
horizon stretching changes the horizontal shape of the barrier but not the
sector dependent conclusion drawn here.

Let
\begin{equation}
    r_+=M+\sqrt{M^2-a^2},
    \qquad
    \Omega_H=\frac{a}{X(r_+)},
    \qquad
    \tilde{\omega}_m\equiv \omega-m\Omega_H .
    \label{eq:horizon_omega_main}
\end{equation}
The shifted frequency $\tilde{\omega}_m=\omega-m\Omega_H$ is the Killing
frequency associated with the horizon-generating vector
$\chi=\partial_t+\Omega_H\partial_\phi$, or equivalently the mode frequency in
the frame co-rotating with the horizon.  The physical ``in'' solution is
chosen to be purely ingoing at the future horizon and to contain incident plus
reflected waves at spatial infinity:
\begin{equation}
    U_{lm}\sim
    \begin{cases}
        A_{\mathrm{in}}U^{\infty,\mathrm{in}}_{\omega lm}
        +A_{\mathrm{out}}U^{\infty,\mathrm{out}}_{\omega lm},
        & r_*\rightarrow+\infty,\\[3pt]
        T_{\omega lm}U^{H,\mathrm{in}}_{\omega lm},
        & r_*\rightarrow-\infty .
    \end{cases}
    \label{eq:boundary_conditions_main}
\end{equation}
This is the usual flux normalization for black hole absorption:
$A_{\mathrm{in}}$, $A_{\mathrm{out}}$, and $T_{\omega lm}$ are the incident,
reflected, and transmitted amplitudes, respectively.  For numerical matching,
one then represents the local ingoing bases by asymptotic series, as in the
standard Kerr and Kerr-Newman treatments~\cite{Futterman1988,Benone2018},
\begin{align}
    U^{\infty,\mathrm{in}}_{\omega lm}(r)
    &=
    \ee^{-\ii\omega r_*}
    \sum_{j=0}^{N_\infty}\frac{c_j^\infty}{r^j},
    && r_*\rightarrow+\infty ,
    \label{eq:infinity_series_main}\\
    U^{H,\mathrm{in}}_{\omega lm}(r)
    &=
    \ee^{-\ii\tilde{\omega}_m r_*}
    \sum_{j=0}^{N_+}c_j^+(r-r_+)^j,
    && r_*\rightarrow-\infty .
    \label{eq:horizon_series_main}
\end{align}
The coefficients $c_j^\infty$ and $c_j^+$ are obtained by requiring
$U^{\infty,\mathrm{in}}_{\omega lm}$ and $U^{H,\mathrm{in}}_{\omega lm}$ to
solve the radial equation order by order in the two asymptotic regions.  Their
overall normalization is a matching convention and is absorbed into the
wave amplitudes.  The outgoing basis at infinity is
$U^{\infty,\mathrm{out}}_{\omega lm}\equiv
\left(U^{\infty,\mathrm{in}}_{\omega lm}\right)^*$ for real $\omega$.

Conservation of the radial Wronskian gives
\begin{equation}
    \left|\frac{A_{\mathrm{out}}}{A_{\mathrm{in}}}\right|^2
    =
    1-
    \frac{\tilde{\omega}_m}{\omega}
    \left|\frac{T_{\omega lm}}{A_{\mathrm{in}}}\right|^2 .
    \label{eq:wronskian_relation_main}
\end{equation}
Therefore modes satisfying $\tilde{\omega}_m<0$, or
$0<\omega<m\Omega_H$ for $m\Omega_H>0$, are superradiantly amplified:
the reflected flux exceeds the incident flux.  This is the usual rotating
black hole superradiance mechanism~\cite{StarobinskyChurilov1973,TeukolskyPress1974}.
In the present notation it appears as a negative signed absorption factor rather than
as a failure of the flux normalization.

The signed absorption factor is defined from the reflection amplitude as
\begin{equation}
    \Gamma_{\omega lm}
    =
    1-\left|\frac{A_{\mathrm{out}}}{A_{\mathrm{in}}}\right|^2
    =
    \frac{\tilde{\omega}_m}{\omega}
    \left|\frac{T_{\omega lm}}{A_{\mathrm{in}}}\right|^2,
    \label{eq:greybody_factor}
\end{equation}
In the superradiant regime $0<\omega<m\Omega_H$, this quantity is a signed
absorption factor: $\Gamma_{\omega lm}<0$ corresponds to superradiant
amplification rather than to a positive transmission probability.  We
therefore also use the amplification factor
\[
    Z_{\mathrm{amp}}
    =
    \left|\frac{A_{\mathrm{out}}}{A_{\mathrm{in}}}\right|^2-1
    =
    -\Gamma_{\omega lm}
\]
when discussing superradiance.
This completes the radial boundary data needed for the absorption problem.

% ======================================================================
\section{Absorption Cross Section and Limiting Regimes}
\label{sec:limiting_behavior}

The observable absorption cross section is assembled from the radial signed
absorption factors and the scalar spheroidal harmonics.  With the standard
plane wave normalization, the partial absorption cross section for an
incidence angle $\gamma$ is
\begin{equation}
    \sigma_{lm}(\omega,\gamma)
    =
    \frac{4\pi^2}{\omega^2}
    \left|S_{\omega lm}(\gamma)\right|^2
    \Gamma_{\omega lm} .
    \label{eq:partial_cross_section}
\end{equation}
The total finite-frequency absorption cross section is then obtained from
the partial wave sum
\begin{equation}
    \sigabs(\omega,\gamma)
    =
    \sum_{l=0}^{\infty}\sum_{m=-l}^{l}
    \sigma_{lm}(\omega,\gamma).
    \label{eq:mode_sum_main}
\end{equation}
For off-axis incidence it is useful to split the mode sum into
co-/counter-rotating azimuthal sectors,
\begin{align}
    \sigma_+(\omega,\gamma)
    &=
    \sum_{l=0}^{\infty}\sum_{m=1}^{l}
    \sigma_{lm}(\omega,\gamma)
    +\frac{1}{2}\sum_{l=0}^{\infty}\sigma_{l0}(\omega,\gamma),
    \label{eq:sigma_plus_def}\\
    \sigma_-(\omega,\gamma)
    &=
    \sum_{l=0}^{\infty}\sum_{m=-l}^{-1}
    \sigma_{lm}(\omega,\gamma)
    +\frac{1}{2}\sum_{l=0}^{\infty}\sigma_{l0}(\omega,\gamma),
    \label{eq:sigma_minus_def}
\end{align}
so that $\sigabs=\sigma_+ + \sigma_-$.  This is a bookkeeping convention rather
than a new observable; the $m=0$ sector is divided equally because it carries
no azimuthal preference.  For a black hole with $a>0$, the $m>0$ modes are the
co-rotating sector in this convention, while $m<0$ modes are counter-rotating.
The physical distinction enters through the radial factor
$\omega-ma/X(r)$ and the horizon frequency
$\tilde{\omega}_m=\omega-m\Omega_H$.  Co-rotating modes can be closer to the
superradiant window and are therefore more strongly affected by rotational
energy exchange, whereas counter-rotating modes are never superradiant for
$\omega>0$ and $a>0$.  The asymmetry between $\sigma_+$ and $\sigma_-$ is thus
a direct finite-frequency probe of frame dragging and of the deformation
dependence of $X(r)$ and $\Omega_H$.

In numerical work the sums are truncated at $L_{\max}$ and increased until the
tail is negligible.  For on-axis incidence only the $m=0$ modes contribute,
whereas equatorial incidence imposes the usual parity selection on the scalar
spheroidal functions.  These selection rules are inherited directly from the
Kerr angular problem; all Johannsen dependence enters through the radial
signed absorption factors.

\subsection{Low-Frequency Limit}
\label{subsec:low_frequency_limit}

In the zero-frequency limit the scalar absorption cross section approaches
the horizon area,
\begin{equation}
    \sigabs(\omega\rightarrow0)\rightarrow A_H .
    \label{eq:low_frequency_area_limit}
\end{equation}
This is the standard area law for a minimally coupled massless scalar in
stationary black hole backgrounds under the usual regularity assumptions on
the zero-frequency solution~\cite{DasGibbonsMathur1997,Higuchi2001,Higuchi2002}.
For the working Johannsen subclass the induced metric on a spatial horizon
section gives
\begin{equation}
    A_H
    =
    \int_0^{2\pi}\!\dd\phi
    \int_0^\pi\!\dd\theta\,
    \sqrt{g_{\theta\theta}g_{\phi\phi}}\bigg|_{r=r_+}
    =
    4\pi X(r_+)
    =
    4\pi\frac{(r_+^2+a^2)A_1(r_+)}{A_2(r_+)} .
    \label{eq:johannsen_horizon_area}
\end{equation}
The last equality uses the KG separability condition
$(r^2+a^2)A_1=X A_2$.  Equation~\eqref{eq:johannsen_horizon_area} makes the
sector dependence transparent: $A_1$ increases or decreases the area through
the numerator, $A_2$ changes it inversely, and $A_5$ has no direct effect
because it only appears in the radial metric component.  The low-frequency
rows of Table~\ref{tab:low_high_frequency_limits} reflect precisely this
structure.  For the positive leading coefficients shown there,
$\alpha_{13}=0.5$ increases $A_H$, $\alpha_{23}=0.5$ decreases it, and
$\alpha_{52}=0.5$ leaves the Kerr area unchanged at every spin.

\subsection{High-Frequency Limit}
\label{subsec:high_frequency_limit}

In the high-frequency limit, the wave problem reduces to the capture of null
geodesics.  For an incidence angle $\gamma$, the geometric cross section is
the area of the critical capture region on the incident wavefront.  Following
the impact plane construction used for Kerr~\cite{Macedo2013}, this area can
be written as
\begin{equation}
    \siggeo(\gamma)
    =
    \frac{1}{2}\int_{-\pi}^{\pi} b_c^2(\chi,\gamma)\,\dd\chi ,
    \label{eq:geometric_cross_section}
\end{equation}
where $\chi$ is the polar angle on the wavefront and $b_c(\chi,\gamma)$ is the
critical impact parameter separating captured and scattered rays.  For each
direction on the wavefront it is determined by the unstable circular orbit
conditions
\begin{equation}
    \mathcal{R}_{\mathrm{geo}}(r_c)=0,
    \qquad
    \frac{\dd\mathcal{R}_{\mathrm{geo}}}{\dd r}(r_c)=0 .
    \label{eq:critical_geodesic_conditions}
\end{equation}
Here $\mathcal{R}_{\mathrm{geo}}$ denotes the radial Hamilton--Jacobi potential
for null geodesics.  The crucial Johannsen feature is that $A_5$ multiplies
the radial kinetic term but does not enter the algebraic conditions
\eqref{eq:critical_geodesic_conditions}.  Hence $A_5$ can modify
finite-frequency radial propagation without moving the geometric capture boundary.

\begin{figure}[H]
    \centering
    \includegraphics[width=0.92\textwidth]{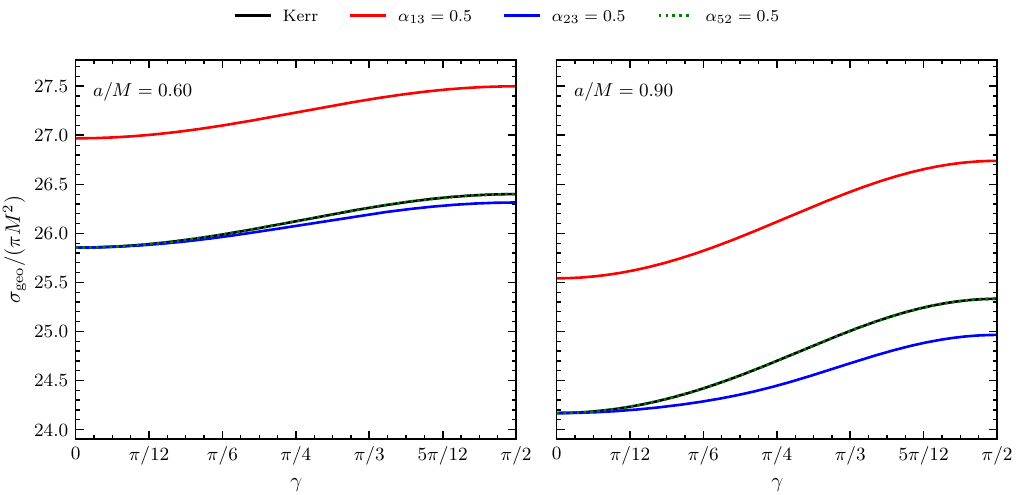}
    \caption{
        High-frequency geometric capture cross section as a function of
        incidence angle for $a/M=0.6$ and $a/M=0.9$.  The $A_5$ deformation is
        degenerate with Kerr in geometric optics, while $A_1$ and
        $A_2$ change the null capture boundary.
    }
    \label{fig:geometric_gamma}
\end{figure}
The incidence-angle scan in Fig.~\ref{fig:geometric_gamma} makes this
separation of sectors explicit.  Comparing the two spin values also makes the
spin dependence visible: increasing the Kerr spin parameter from $a/M=0.6$ to
$a/M=0.9$ lowers the overall capture area for the displayed families.  This
decrease reflects the shrinkage and distortion of the critical null capture
region at higher rotation, rather than a uniform rescaling of the impact
plane.  The $A_1$ sector changes the effective radial size $X(r)$ and
therefore shifts the capture area already for on-axis incidence.  The $A_2$
sector is more selective: it can be degenerate with Kerr in the on-axis
entries of Table~\ref{tab:low_high_frequency_limits}, but its deviation grows
with incidence angle, and the departure is more pronounced at $a/M=0.9$ than
at $a/M=0.6$.  This reflects the fact that the $A_2$ deformation enters the
spin-coupled azimuthal part of the null capture conditions.  By contrast, the
$A_5$ curve lies on top of Kerr for both displayed spins, confirming its
invisibility to the leading null capture observable.  This does not mean that
$A_5$ is absent from the spacetime geometry; it remains part of the radial
metric component.  The equality is also visible in
Table~\ref{tab:low_high_frequency_limits}: the on-axis high-frequency entries
for $\alpha_{52}=0.5$ are identical to Kerr for all listed spins.

The approach to $\siggeo$ is not monotonic.  Schwarzschild calculations by
Sanchez showed that the high-frequency absorption cross section oscillates
around the geometric limit~\cite{Sanchez1978}.  Complex angular momentum
analyses later related this oscillatory pattern to Regge poles and to the
orbital frequency and Lyapunov exponent of unstable null
geodesics~\cite{Decanini2011,DecaniniFine2011}.  For a static spherically symmetric
black hole with critical impact parameter $b_c$, orbital frequency $\Omega$,
Lyapunov exponent $\Lambda$, and $\beta=\Lambda/\Omega$, the leading sinc
approximation can be written as
\begin{equation}
    \frac{\sigabs(\omega)}{\siggeo}
    \simeq
    1
    -8\pi\beta\,\ee^{-\pi\beta}
    \sinc\!\left(\frac{2\pi\omega}{\Omega}\right),
    \qquad
    \sinc x\equiv\frac{\sin x}{x}.
    \label{eq:spherical_sinc_approx}
\end{equation}
For Schwarzschild, $\Omega=1/b_c$ and $\beta=1$.  For on-axis incidence on a
rotating black hole, the same physical idea survives but the polar photon
orbit has $b_c\neq1/\Omega$.  Motivated by the Kerr on-axis approximation of
Macedo et al.~\cite{Macedo2013}, we use the following Johannsen polar-orbit
analogue as a qualitative high-frequency guide:
\begin{equation}
    \frac{\sigabs(\omega)}{\siggeo}
    \simeq
    1
    -
    \frac{8\pi\beta\,\ee^{-\pi\beta}}{\Omega^2 b_c^2}
    \sinc\!\left(\frac{2\pi\omega}{\Omega}\right),
    \qquad
    \beta=\frac{\Lambda}{\Omega},
    \label{eq:rotating_sinc_approx}
\end{equation}
where $b_c$, $\Omega$, and $\Lambda$ are computed from the unstable polar null
orbit of the Johannsen geometry.  In the present work this approximation is
used as a high-frequency guide rather than as
a replacement for the finite-frequency mode sum.  It clarifies which part of
the oscillatory absorption spectrum is controlled by null orbit data and which
part must still be attributed to the full radial wave propagation.
The dotted curves shown with the on-axis spectra below are generated from
Eq.~\eqref{eq:rotating_sinc_approx}.

\begin{table}[H]
    \centering
    \caption{
        Low- and high-frequency limits of the scalar absorption cross section.
        The low-frequency limit is the horizon area, while the high-frequency
        limit is the on-axis geometric capture cross section.  All values are
        given in units of $\pi M^2$.
    }
    \label{tab:low_high_frequency_limits}
    \scriptsize
    \setlength{\tabcolsep}{4pt}
    \begin{tabular}{llccccc}
        \toprule
        Model & Limit
        & $a/M=0$
        & $a/M=0.3$
        & $a/M=0.6$
        & $a/M=0.9$
        & $a/M=0.99$ \\
        \midrule
        Kerr
        & $\sigma(\omega\simeq0)$
        & 16.000 & 15.632 & 14.400 & 11.487 & 9.129 \\
        & $\sigma(\omega M\gg1)$
        & 27.000 & 26.726 & 25.855 & 24.168 & 23.409 \\
        \midrule
        $\alpha_{13}=0.5$
        & $\sigma(\omega\simeq0)$
        & 17.000 & 16.679 & 15.635 & 13.427 & 12.201 \\
        & $\sigma(\omega M\gg1)$
        & 27.982 & 27.738 & 26.969 & 25.542 & 24.937 \\
        \midrule
        $\alpha_{23}=0.5$
        & $\sigma(\omega\simeq0)$
        & 15.059 & 14.650 & 13.263 & 9.827 & 6.830 \\
        & $\sigma(\omega M\gg1)$
        & 27.000 & 26.726 & 25.855 & 24.168 & 23.409 \\
        \midrule
        $\alpha_{52}=0.5$
        & $\sigma(\omega\simeq0)$
        & 16.000 & 15.632 & 14.400 & 11.487 & 9.129 \\
        & $\sigma(\omega M\gg1)$
        & 27.000 & 26.726 & 25.855 & 24.168 & 23.409 \\
        \bottomrule
    \end{tabular}
\end{table}

% ======================================================================
\section{Finite-Frequency Absorption Results}
\label{sec:finite_frequency_results}

The numerical calculation follows the partial wave construction in
Sec.~\ref{sec:limiting_behavior}.  For each $(\omega,l,m)$ we first compute
the scalar spheroidal eigenvalue, then integrate the Johannsen radial equation
from the near horizon region to the asymptotic matching region, and finally
assemble the cross section from the signed absorption factors.  For each relative
deviation and root-mean-square (RMS) diagnostic, the Kerr reference spectrum
is recomputed on the same frequency grid rather than interpolated from a
separate dataset.  The radial integration starts from the ingoing
horizon expansion and is matched to independent ingoing and outgoing
asymptotic solutions at large radius.  The initial-value problem is solved in
the Boyer--Lindquist-like radius using an adaptive high-order Runge--Kutta
method (DOP853), with production tolerances typically set to
$\mathrm{rtol}=10^{-9}$ and $\mathrm{atol}=10^{-11}$ and tightened in targeted
reruns when matching diagnostics require it.  Expansion orders and matching
radii are raised until the extracted absorption factor is stable.  On-axis
spectra use the exact $m=0$ selection rule, whereas off-axis spectra use the
full triangular mode set $0\le l\le L_{\max}$ and $-l\le m\le l$.  In all
production plots the mode cutoff is increased until the pointwise
cross section sum is trusted, with targeted radial reruns used only for
isolated points that fail the matching or truncation checks.

Before discussing the spectra, we quantify why the leading Johannsen
coefficients are used as the main deformation parameters.  For one active
coefficient of radial falloff order $n$, define the pointwise relative
deviation and its RMS average over a fixed frequency grid by
\begin{align}
    \delta_n(\omega_j)
    &=
    \frac{\sigma_n(\omega_j)-\sigma_{\mathrm{Kerr}}(\omega_j)}
         {\sigma_{\mathrm{Kerr}}(\omega_j)},
    \label{eq:delta_n_hierarchy}\\
    S_n
    &=
    \left[
        \frac{1}{N_\omega}\sum_{j=1}^{N_\omega}
        \delta_n(\omega_j)^2
    \right]^{1/2}.
    \label{eq:Sn_hierarchy}
\end{align}
Only points for which both the Johannsen numerator and the Kerr denominator
pass the trust checks enter $S_n$.

\begin{figure}[H]
    \centering
    \includegraphics[width=0.92\textwidth]{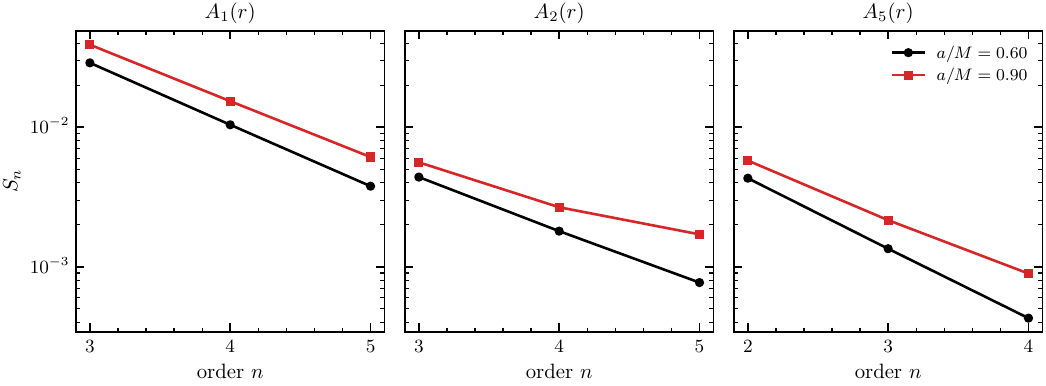}
    \caption{
        Hierarchy of Johannsen coefficients with increasing radial falloff.  The plotted
        quantity is the RMS sensitivity $S_n$ defined by
        Eqs.~\eqref{eq:delta_n_hierarchy} and \eqref{eq:Sn_hierarchy}, with
        $M=1$, $\gamma=0$, coefficient amplitude $p=0.3$,
        $\omega M\in[0.1,1.5]$, and $N_\omega=101$.  The Kerr denominators are
        recomputed on the same grid for each spin.  The panels correspond to
        the $A_1$, $A_2$, and $A_5$ sectors.
    }
    \label{fig:hierarchy_rms_main}
\end{figure}

The trend extracted in Fig.~\ref{fig:hierarchy_rms_main} is monotonic in all
sampled sectors: the absorption response decreases as the radial falloff order
is raised.  This does not prove that every possible higher order deformation is
negligible, but it supports a controlled leading coefficient study: for the
same asymptotic coefficient amplitude, the lowest allowed powers give the
largest finite-frequency response.  The frequency resolved curves underlying
this RMS compression are given in Appendix~\ref{app:numerical_convergence}.

The validation strategy is tied to the limiting results of
Sec.~\ref{sec:limiting_behavior}.  The same code recovers the Kerr limit,
checks $\sigma=\sigma^+ +\sigma^-$ whenever a co-/counter-rotating split is computed,
verifies the low-frequency approach to the horizon area, and compares the
high-frequency envelope with the geometric capture values in
Table~\ref{tab:low_high_frequency_limits}.  These tests are important because
the finite-frequency differences discussed below are typically smaller than
the gross variation of the total cross section with spin.

\FloatBarrier

\subsection{Leading-Order On-Axis Spectra}
\label{subsec:onaxis_absolute_results}

\begin{figure}[H]
    \centering
    \begin{subfigure}{0.48\textwidth}
        \centering
        \includegraphics[width=\linewidth]{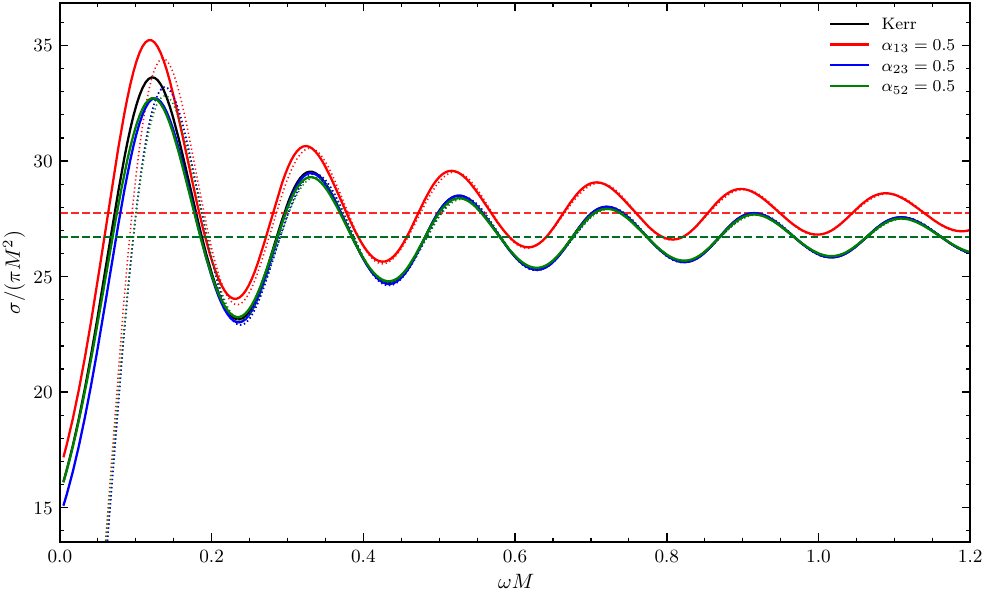}
        \caption{$a/M=0.3$}
    \end{subfigure}
    \hfill
    \begin{subfigure}{0.48\textwidth}
        \centering
        \includegraphics[width=\linewidth]{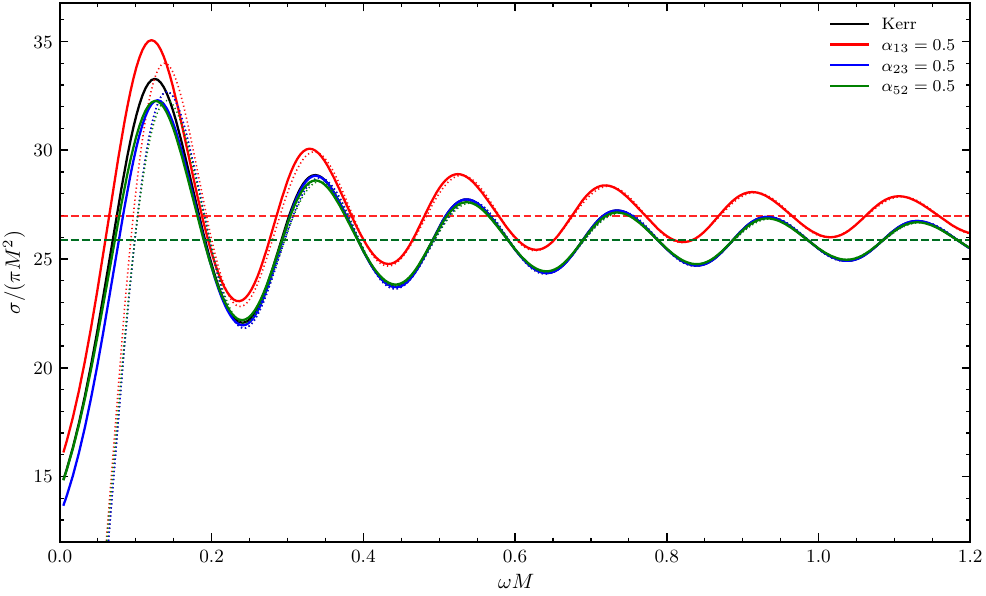}
        \caption{$a/M=0.6$}
    \end{subfigure}

    \vspace{0.6em}
    \begin{subfigure}{0.48\textwidth}
        \centering
        \includegraphics[width=\linewidth]{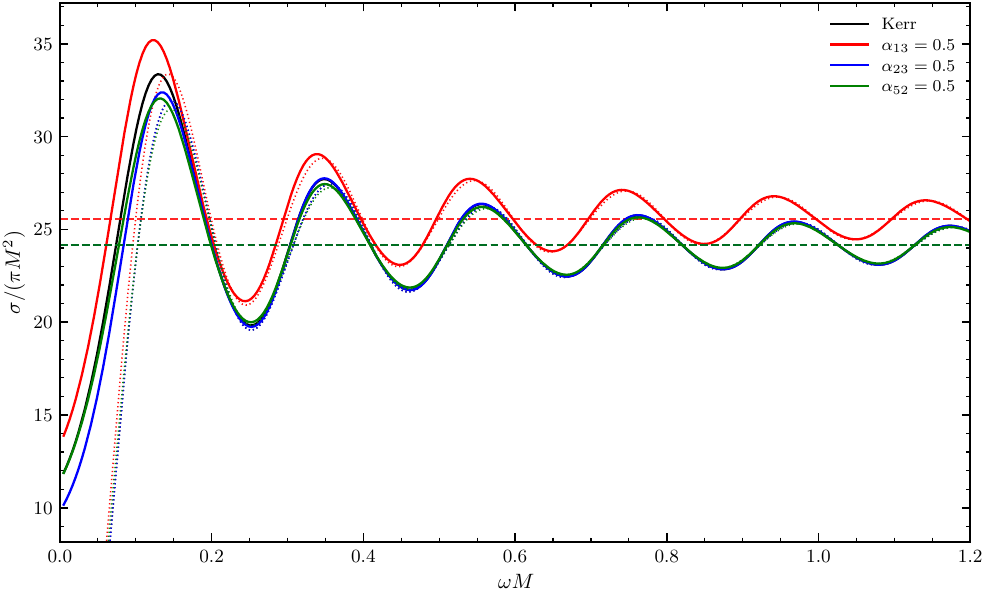}
        \caption{$a/M=0.9$}
    \end{subfigure}
    \hfill
    \begin{subfigure}{0.48\textwidth}
        \centering
        \includegraphics[width=\linewidth]{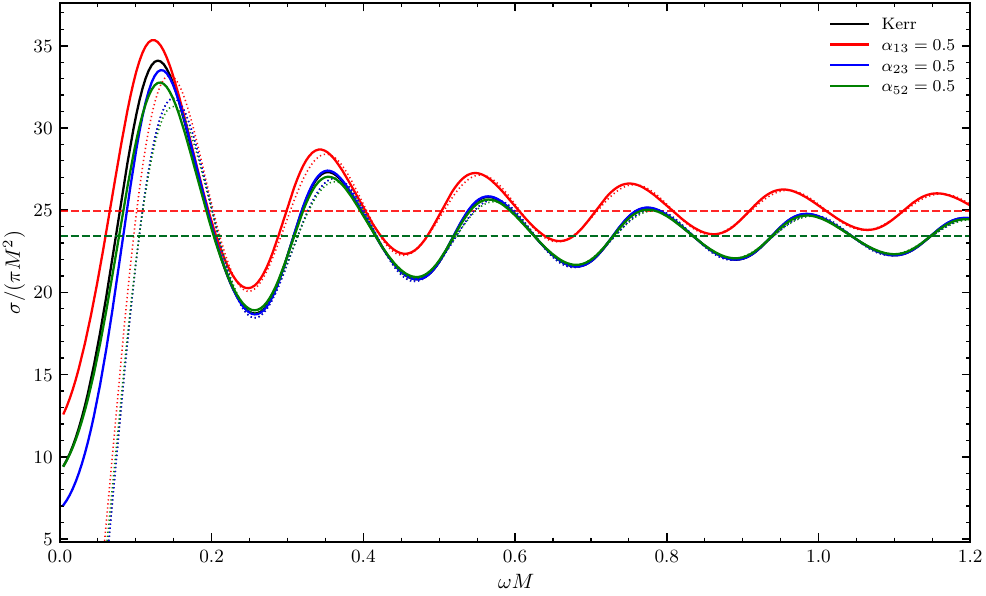}
        \caption{$a/M=0.99$}
    \end{subfigure}
    \caption{
        On-axis total absorption spectra for Kerr and for the leading
        Johannsen deformations with coefficient amplitude $0.5$.  The dashed
        horizontal lines show the corresponding high-frequency
        geometric capture limits, and the dotted curves show the associated
        high-frequency sinc estimates from
        Eq.~\eqref{eq:rotating_sinc_approx}.
    }
    \label{fig:onaxis_spectra_spin_grid}
\end{figure}

We first consider on-axis incidence, $\gamma=0$, for which only the $m=0$
spheroidal harmonics contribute to the incident plane wave.  This setup is a
clean baseline because it removes the azimuthal mixing associated with
off-axis incidence while retaining the full radial boundary-value problem.  The
spectra in Fig.~\ref{fig:onaxis_spectra_spin_grid} compare Kerr with the
three single parameter Johannsen families
$\alpha_{13}=0.5$, $\alpha_{23}=0.5$, and $\alpha_{52}=0.5$.

The low-frequency ordering follows the area formula in
Eq.~\eqref{eq:johannsen_horizon_area}: positive $\alpha_{13}$ increases
$A_H$, positive $\alpha_{23}$ decreases it through the denominator
$A_2(r_+)$, and positive $\alpha_{52}$ leaves it unchanged.  The same pattern
is visible in the first rows of Table~\ref{tab:low_high_frequency_limits} and
in the left edge of the spectra.  In the high-frequency regime the curves approach the
geometric reference lines discussed in Sec.~\ref{subsec:high_frequency_limit}.
The $A_5$ curves again share the Kerr limiting value, while the
finite-frequency part of the spectrum is shifted because $A_5$ enters the
radial kinetic operator and the tortoise coordinate.  This is the first
numerical indication that finite-frequency absorption probes more structure
than the two limiting cross sections.

To isolate the sector dependence, we use the relative deviation
\begin{equation}
    \frac{\delta\sigma}{\sigma_{\mathrm{Kerr}}}
    =
    \frac{\sigma_{\mathrm{J}}-\sigma_{\mathrm{Kerr}}}
         {\sigma_{\mathrm{Kerr}}}.
    \label{eq:relative_deviation}
\end{equation}
This normalization removes the common Kerr envelope and makes the sign,
frequency location, and oscillatory phase of each deformation easier to
compare.

\begin{figure}[H]
    \centering
    \includegraphics[width=0.96\textwidth]{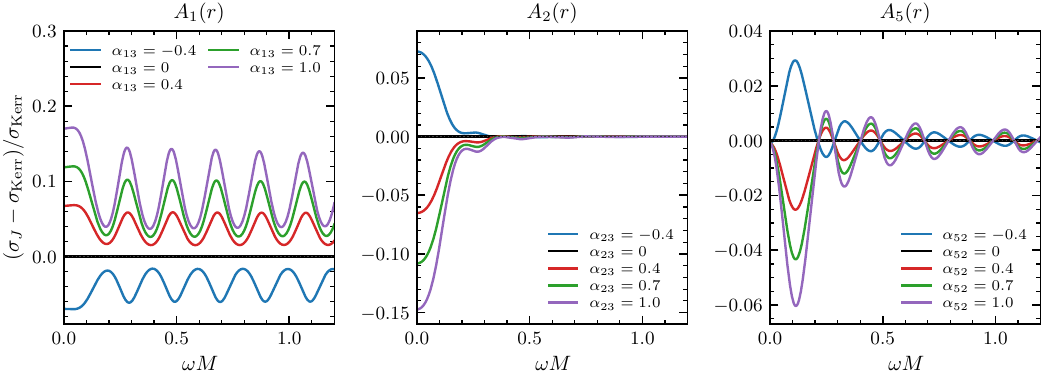}
    \caption{
        Relative deviations from Kerr by deformation sector for on-axis incidence
        at fixed spin $a/M=0.6$ and $\gamma=0$.  From left to right the panels
        show the $A_1$, $A_2$, and $A_5$ sectors; in each panel only the
        corresponding leading Johannsen parameter is varied.
    }
    \label{fig:phase7_relative_scans}
\end{figure}

The relative spectra in Fig.~\ref{fig:phase7_relative_scans} make clear that
the three metric sectors do not act as interchangeable rescalings of the Kerr
spectrum.  The deviations are not merely smooth offsets: the $A_1$ and $A_5$
sectors in particular show oscillatory structure across the finite-frequency
range.  This behavior reflects changes in the radial barrier and in the
relative phase of the partial-wave absorption pattern, and the same
frequency-resolved behavior is visible in the hierarchy curves shown in
Fig.~\ref{fig:hierarchy_relative_appendix}.  The $A_1$ and $A_2$ panels carry
the expected imprint of the low- and high-frequency limits, because these
functions enter $X(r)$ and hence both $A_H$ and the geometric capture problem.
The $A_5$ sector is qualitatively different.  Its geometric cross section
coincides with Kerr for the full incidence-angle range shown in
Fig.~\ref{fig:geometric_gamma}, yet its finite-frequency relative deviation is
nonzero.  The comparison should not be read as a point-by-point comparison
between an on-axis wave spectrum and an off-axis geometric scan.  Rather, it
identifies the mechanism: $A_5$ is absent from the leading null capture
boundary but present in the radial wave propagation.  It is therefore
geometry-invisible only in the restricted sense of the leading null capture
observable, but wave-visible at finite frequency.  The
corresponding total absorption curves are shown in
Appendix~\ref{app:numerical_convergence}.

\FloatBarrier

\subsection{Angular Dependence and Azimuthal Asymmetry}
\label{subsec:offaxis_asymmetry_results}

Off-axis incidence activates the full set of azimuthal modes and tests whether
the conclusions above are artifacts of the on-axis $m=0$ selection rule.  We
use the co-/counter-rotating split in Eqs.~\eqref{eq:sigma_plus_def} and
\eqref{eq:sigma_minus_def} to define
\begin{equation}
    \mathcal{A}(\omega)
    =
    \frac{\sigma^-(\omega)-\sigma^+(\omega)}
         {\sigma^-(\omega)+\sigma^+(\omega)} .
    \label{eq:pm_asymmetry_def}
\end{equation}
For the sign convention used here, $\mathcal{A}>0$ means that the
counter-rotating sector contributes more strongly to the absorption sum,
whereas $\mathcal{A}<0$ means that the co-rotating sector dominates.

\begin{figure}[H]
    \centering
    \includegraphics[width=0.86\textwidth]{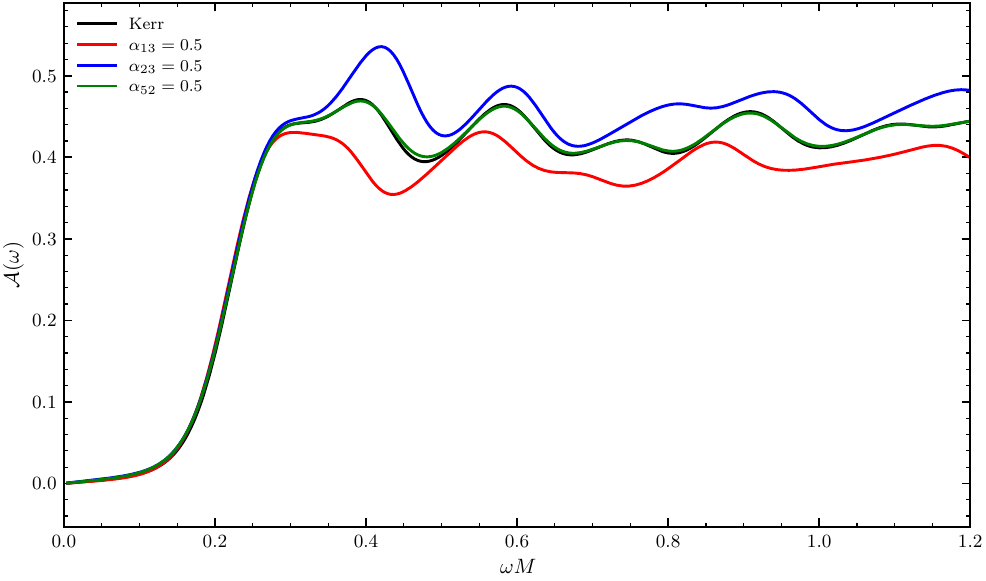}
    \caption{
        Azimuthal asymmetry for off-axis incidence at $a/M=0.9$ and $\gamma=\pi/3$,
        measured by $\mathcal{A}(\omega)$ in
        Eq.~\eqref{eq:pm_asymmetry_def}.  The calculation uses the full
        triangular mode set rather than the on-axis $m=0$ selection rule.
    }
    \label{fig:pm_asymmetry}
\end{figure}

In Fig.~\ref{fig:pm_asymmetry}, the off-axis calculation produces a clear
frequency dependent imbalance between the two azimuthal sectors.  This
imbalance is expected for rotating holes: positive and negative $m$ modes
experience different frame dragging shifts in $\omega-ma/X(r)$ and different
horizon frequencies $\tilde{\omega}_m$.  Johannsen deformations modify this
imbalance in a sector dependent way.  $A_1$ and $A_2$ change $X(r)$ and
$\Omega_H$ directly, while $A_5$ modifies the radial transmission problem
without changing the geometric capture boundary.  The $A_5$ curve is therefore
close to Kerr but not identical, which is another finite-frequency imprint of
the radial kinetic sector.  The result confirms that the finite-frequency
signal is not restricted to the on-axis special case.  A
direct plot of $\sigma^+$ and $\sigma^-$ is included in
Appendix~\ref{app:numerical_convergence} to show that
Fig.~\ref{fig:pm_asymmetry} is not produced by a denominator artifact.

\FloatBarrier

\subsection{Single-Mode Superradiance}
\label{subsec:superradiant_mode_results}

The signed absorption factor in Eq.~\eqref{eq:greybody_factor} also carries the
superradiant information.  In the usual rotating black hole window
$0<\omega<m\Omega_H$, the reflected flux is larger than the incident flux and
$\Gamma_{\omega lm}<0$
\cite{StarobinskyChurilov1973,TeukolskyPress1974}.  Through
Eq.~\eqref{eq:partial_cross_section}, the corresponding single-mode
contribution to the plane-wave absorption cross section becomes negative in
this interval.  We therefore plot the $l=m=1$ partial absorption cross section
at $a/M=0.99$ and $\gamma=\pi/2$, using an inset to resolve the narrow
negative band while keeping the finite-frequency scale of the mode visible in
the main panel.  An equivalent positive amplification factor,
\begin{equation}
    Z_{\mathrm{amp}}(\omega)
    =
    \left|\frac{A_{\mathrm{out}}}{A_{\mathrm{in}}}\right|^2-1
    =
    -\Gamma_{\omega lm},
    \qquad 0<\omega<m\Omega_H ,
    \label{eq:superradiant_amplification_factor}
\end{equation}
is displayed separately in Appendix~\ref{app:numerical_convergence}.

\begin{figure}[H]
    \centering
    \includegraphics[width=0.92\textwidth]{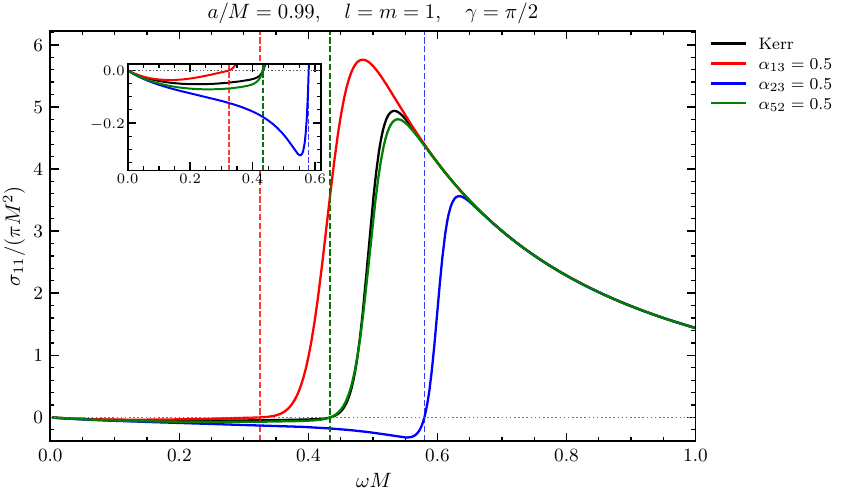}
    \caption{
        Partial absorption cross section $\sigma_{11}/(\pi M^2)$ for the
        $l=m=1$ scalar mode at $a/M=0.99$ and $\gamma=\pi/2$.  The main panel
        shows the finite-frequency scale of this mode, while the inset
        magnifies the low-frequency superradiant regime, where
        $\Gamma_{\omega 11}<0$ and hence $\sigma_{11}<0$.  Negative values in
        the inset are a mode-level superradiant effect and do not imply a
        negative total absorption cross section.  The vertical dashed lines
        mark the thresholds $\omega=m\Omega_H$ for the corresponding cases.
    }
    \label{fig:superradiance_zoom}
\end{figure}

Figure~\ref{fig:superradiance_zoom} should be read as a mode-resolved partial
absorption plot rather than as a total cross section.  The main panel shows
that the same $l=m=1$ contribution becomes positive and grows outside the
superradiant band, whereas the inset isolates the negative low-frequency
portion.  The horizon angular velocity is $\Omega_H=a/X(r_+)$, so any
deformation that changes $X(r_+)$ also changes the edge of the superradiant
band.  For the positive coefficients shown here, $\alpha_{13}=0.5$ increases
$X(r_+)$ and shifts the threshold to smaller $\omega M$, whereas
$\alpha_{23}=0.5$ decreases $X(r_+)$ and pushes the threshold upward.  A pure
$\alpha_{52}$ deformation leaves $X(r_+)$, and therefore $\Omega_H$, equal to
the Kerr value, but it still affects the detailed radial transmission through
the radial kinetic term.  The superradiant mode therefore gives another
example of the same sector separation found in the absorption spectra: the
threshold is controlled by the horizon value of $X(r)$, while the shape and
depth of the partial-cross-section curve are sensitive to the full wave
operator.

\FloatBarrier

% ======================================================================
\section{Discussion and Conclusion}
\label{sec:discussion}

We have constructed the scalar absorption problem for a Klein–Gordon-separable Johannsen subclass and traced the leading deformation sectors through their analytic limits and the finite-frequency 
partial-wave calculation. The separability condition ties the Johannsen radial functions through
$X(r)=(r^2+a^2)A_1/A_2$, leaving three independent sectors in the simplified
PPN compatible model: $A_1$, $A_2$, and $A_5$.  This structure is physically instructive because it cleanly separates metric components that govern the horizon area and null-capture geometry from 
those that modify only the radial wave propagation.

Our central result is that finite-frequency wave absorption resolves a fundamental degeneracy inherent to purely geometric probes. The functions $A_1$ and $A_2$
change the radial size function $X(r)$ and therefore appear in the
low-frequency area law, in the high-frequency capture cross section, and in
finite-frequency spectra.  In contrast, $A_5$ changes the radial kinetic
operator and the tortoise coordinate, but does not change the horizon area or
the null geodesic capture boundary for a pure $A_5$ deformation.  The
finite-frequency relative deviation curves nevertheless show a nonzero
response to $\alpha_{52}$.  This constitutes the primary novel finding of our calculation: the $A_5$ sector is invisible to both the low-frequency area law and the leading high-frequency 
null-capture observable, yet it becomes manifest once wave-optics effects are included at finite frequencies. In other words, scalar wave absorption provides a diagnostic of radial propagation 
sectors that remain entirely degenerate in geometric-optics observables.

  This wave-mechanical discrimination carries direct significance for strong-field tests of the no-hair theorem. Because geometric probes such as shadow morphology and null-capture cross sections 
  are insensitive to pure $A_5$-type deformations, they cannot alone exhaust the space of possible Kerr deviations. Our results demonstrate that finite-frequency absorption cross sections can break 
  this degeneracy, offering a complementary probe of metric deformations that geometric methods miss. Consequently, wave-optics observables—accessible through black hole absorption, quasinormal mode 
  spectroscopy, and related scattering phenomena—are essential for a complete and systematic test of whether astrophysical black holes are uniquely described by the Kerr geometry, or whether they 
  harbor deviations permitted by more general stationary, axisymmetric spacetimes.

This perspective is consistent with recent full-wave studies of regular
black-hole backgrounds.  In the Frolov case, the absorption and scattering
spectra of Frolov, Reissner--Nordstr{\"o}m, and Hayward geometries were found
to become nearly indistinguishable once the relevant critical or glory impact
parameters were matched, emphasizing the leading role of photon-sphere data
in the intermediate-to-high-frequency regime~\cite{TangHuangZhang2026Frolov}.
The Johannsen calculation here addresses a complementary question: rather
than matching different static geometries by their photon-sphere scales, we
isolate a rotating deformation sector, $A_5$, that is invisible to the
leading null-capture observable but remains visible in the finite-frequency
radial wave problem.

The numerical hierarchy test provides a consistency check on this interpretation: at fixed asymptotic coefficient amplitude, the lowest allowed powers produce the largest finite-frequency response, 
supporting the use of $\alpha_{13}$, $\alpha_{23}$, and $\alpha_{52}$ as the primary deformation parameters. Furthermore, the off-axis calculation, the co-/counter-rotating asymmetry, and the 
single-mode superradiant threshold all confirm that the sector-dependent signal is not an artifact of the on-axis $m=0$ selection rule. In particular, $A_1$ and $A_2$ shift $X(r_+)$ and therefore 
the horizon angular velocity $\Omega_H=a/X(r_+)$, moving the superradiant threshold, whereas $A_5$ leaves
the threshold location unchanged but still modifies the radial amplification profile—another manifestation of the same wave-visible/geometric-invisible separation.

Several qualifications should be kept explicit. The calculation is performed for a test, massless scalar field, so it does not include backreaction, massive field bound states, or spin-dependent 
couplings. The Johannsen functions are treated in a truncated leading-coefficient model; this is motivated by the hierarchy test, but it is not a proof that every higher-order combination is 
observationally negligible. Finally, the mapping between finite Johannsen truncations and exact alternatives to Kerr should be interpreted as a weak-field or sector-level comparison rather than a 
unique global identification.

Our framework can be extended in several directions. The most direct next step is a broader parameter survey, including incidence-angle dependence, larger spin grids, and systematic comparisons with 
exact separable metrics such as Kerr–Newman or Kerr–Sen where available. A second direction is to move beyond massless scalar waves, either by including a field mass or by studying electromagnetic 
and gravitational perturbations. More broadly, the method developed here can be carried to more general rotating black holes, including parametrized or specific theoretical geometries whose 
separability properties differ from the Johannsen subclass. In that setting, absorption, shadows, quasinormal mode spectra, and superradiant amplification could be compared within a unified 
strong-field test: geometric observables constrain the capture sector, while finite-frequency waves probe radial propagation that geometric optics can miss—thereby furnishing a comprehensive, 
wave-mechanics-based diagnostic of deviations from the Kerr metric and, by extension, from the black-hole uniqueness theorems.

\section*{ACKNOWLEDGMENTS}

This work is supported by the National Natural Science Foundation of China
(NSFC) under Grant nos.~12275106, ~12235019 and Shandong Provincial Natural
Science Foundation under grant No.~ZR2024QA032.

% ======================================================================
\clearpage
\appendix

\section{Additional Numerical Results}
\label{app:numerical_convergence}

This appendix collects the numerical curves referenced in the main text but
not displayed there: the tortoise-coordinate view of the effective potential,
the frequency resolved hierarchy behind the RMS measure,
the total spectra behind the relative deviations, and the direct
$\sigma^\pm$ split behind the azimuthal asymmetry.  It also shows the
positive amplification form of the superradiant mode discussed in
Sec.~\ref{subsec:superradiant_mode_results}.

\begin{figure}[H]
    \centering
    \includegraphics[width=0.62\textwidth]{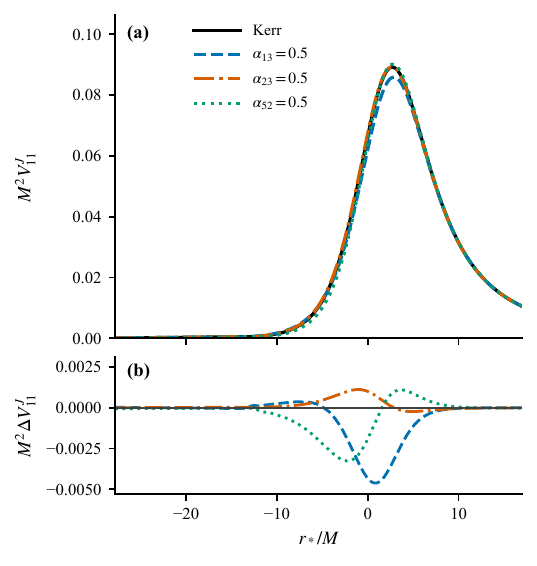}
    \caption{
        Dimensionless scalar wave effective potential $M^2V^J_{11}$ plotted
        against the tortoise coordinate $r_*/M$ for the same parameter choices
        as Fig.~\ref{fig:leading_effective_potential}: $a/M=0.6$,
        $M\omega=0.3$, and $(l,m)=(1,1)$.  The lower panel shows the
        difference from Kerr after interpolation onto a common $r_*/M$ grid.
    }
    \label{fig:leading_effective_potential_rstar}
\end{figure}

\begin{figure}[H]
    \centering
    \includegraphics[width=0.92\textwidth]{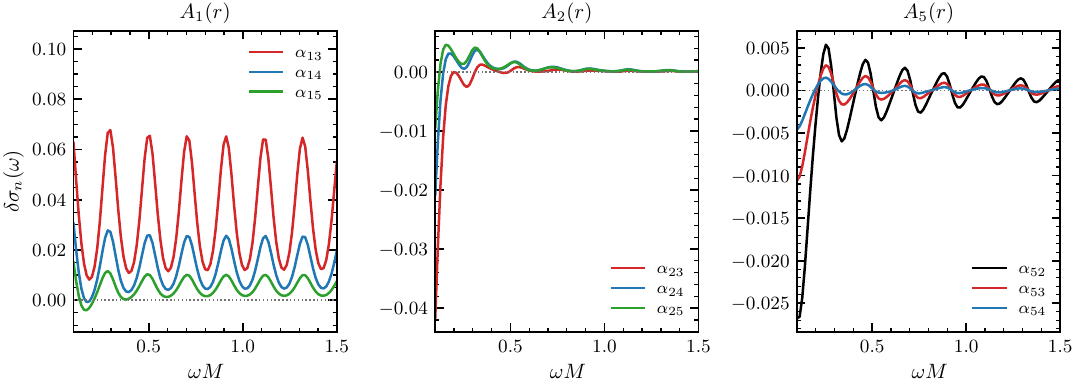}
    \caption{
        Frequency dependence of the relative deviations used in the hierarchy
        comparison at $a/M=0.9$.  The three panels correspond to the $A_1$,
        $A_2$, and $A_5$ sectors and show how the signed deviations vary before
        they are compressed into the RMS quantity $S_n$ in
        Fig.~\ref{fig:hierarchy_rms_main}.
    }
    \label{fig:hierarchy_relative_appendix}
\end{figure}

\begin{figure}[H]
    \centering
    \includegraphics[width=0.92\textwidth]{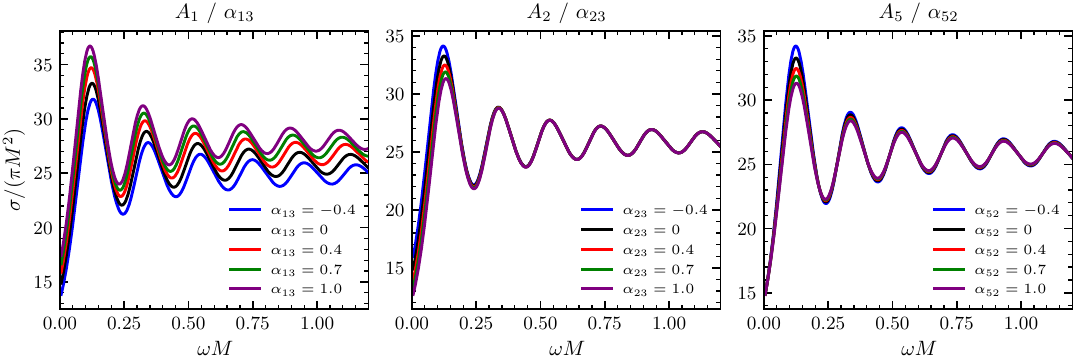}
    \caption{
        Total absorption spectra for the same on-axis setup as
        Fig.~\ref{fig:phase7_relative_scans}.  The curves give the absolute
        cross sections from which the relative deviations are formed and make
        the common Kerr envelope visible before normalization.
    }
    \label{fig:phase7_total_absorption_appendix}
\end{figure}

\begin{figure}[H]
    \centering
    \includegraphics[width=0.86\textwidth]{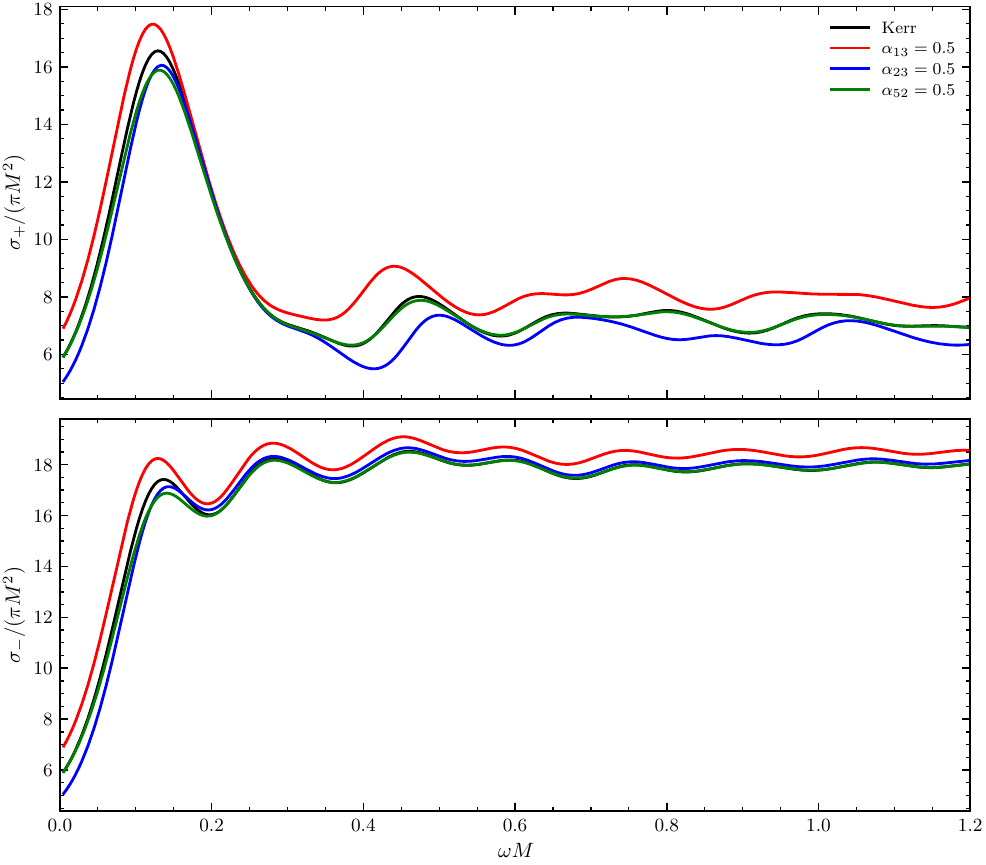}
    \caption{
        Co- and counter-rotating azimuthal contributions to the absorption
        spectrum for the off-axis incidence case used in
        Fig.~\ref{fig:pm_asymmetry}.  The split shows the separate
        $\sigma^+$ and $\sigma^-$ curves whose imbalance defines
        $\mathcal{A}(\omega)$.
    }
    \label{fig:sigma_plus_minus_diagnostic}
\end{figure}

\begin{figure}[H]
    \centering
    \includegraphics[width=0.86\textwidth]{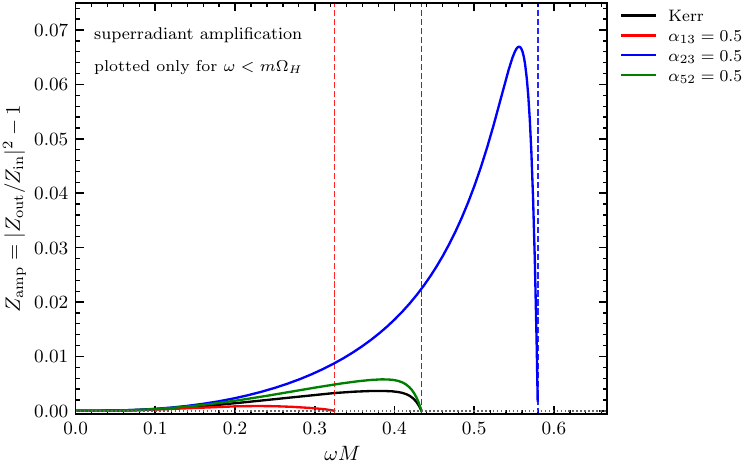}
    \caption{
        Positive amplification factor of the $l=m=1$ superradiant mode at
        $a/M=0.99$, matching the spin used in
        Fig.~\ref{fig:superradiance_zoom}.  The plotted quantity is
        $Z_{\mathrm{amp}}=|A_{\mathrm{out}}/A_{\mathrm{in}}|^2-1
        =-\Gamma_{\omega 11}$ inside the superradiant interval.  Vertical
        dashed lines mark $\omega=m\Omega_H$ for each metric.
    }
    \label{fig:superradiance_amplification_appendix}
\end{figure}

\clearpage

\bibliographystyle{unsrtnat}
\bibliography{johannsenNotes}

\end{document}